# Cellular automata inspired multistable origami metamaterials for mechanical learning

*Zuolin Liu, Hongbin Fang\*, Jian Xu, and K.W. Wang*

Zuolin Liu, Hongbin Fang, Jian Xu

Institute of AI and Robotics, State Key Laboratory of Medical Neurobiology, MOE Engineering Research Center of AI & Robotics, Fudan University, Shanghai 200433, China

E-mail: fanghongbin@fudan.edu.cn

Zuolin Liu, K.W. Wang

Department of Mechanical Engineering, University of Michigan, Ann Arbor, MI 48109, USA

**Abstract**: Recent advances in multistable metamaterials reveal a link between structural configuration transition and Boolean logic, heralding a new generation of computationally capable intelligent materials. To enable higher-level computation, existing computational frameworks require the integration of large-scale networked logic gates, which places demanding requirements on the fabrication of materials counterparts and the propagation of signals. Inspired by cellular automata, we propose a novel computational framework based on multistable origami metamaterials by incorporating reservoir computing, which can accomplish high-level computation tasks without the need to construct a logic gate network. This approach thus eleimates the demanding requirements for fabrication of materials and signal propagation when constructing large-scale networks for high-level computation in conventional mechano-logic. Using the multistable stacked Miura-origami metamaterial as a validation platform, digit recognition is successfully implemented through experiments by a single actuator. Moreover, complex tasks, such as handwriting recognition and 5-bit memory tasks, are also shown to be feasible with the new computation framework. Our research represents a significant advancement in developing a new generation of intelligent materials with advanced computational capabilities. With continued research and development, these materials could have a transformative impact on a wide range of fields, from computational science to material mechano-intelligence technology and beyond.

# 1. Introduction

Material properties such as shape reconfigurability[1], programmable piecewise stiffness[2], multistability[3], and reprogrammable modulus[4] have long been sought by researchers for functional flexibility. Today, a revolutionary idea of incorporating computing power into materials allows the integration of mechanical properties with information processing. By enabling materials to interact with their external environment and perceive and process information through autonomous shape morphing, intelligence emerges in mechanics. Many examples in nature, such as the Venus flytrap[5–7] and mollusks[8], also provide evidence for the incorporation of computational capability into a physical body.

For engineering materials, abstracting mechanical bits from multistable structural configurations[9–13] has been recognized as a tenable approach to achieving mechanical computing. The distinct stable configurations endow multistable metamaterials with nonvolatile mechanical memory[13]. Moreover, the write, erase, and rewrite operations could be efficiently achieved by exploiting the snap-through transitions among different stable configurations. By incorporating multiple cells and by tailoring the mechanical coupling between them, the transition sequences become customizable, thus enabling input-output mapping similar to some basic logic gates[9]. With more mechanical bits being integrated, complex calculations, such as full adders and logic circuits[11,14], have also been achieved. Particularly, higher-level computing or programmable logic gates were explored[15,16] but usually require scalable integrated circuits embedded into mechanical metamaterials. Note that the fundamental principle of mechanical computing follows the framework of modern electric computation, hence, high-level computing operations can be implemented by integrating networked basic logic gates. Based on the one-to-one correspondence between the material components and electronic elements, high-level mechanical computing will encounter bottlenecks due to the explosive growth in demand for material components and the strict requirements for signal propagation efficiency in materials.

Inspired by Cellular Automata (CA)[17], discrete computing paradigms with Turing completeness[18], this paper proposes a novel computational framework based on multistable origami metamaterials to address the above-mentioned bottlenecks in achieving high-level mechanical computation. The framework, by treating the metamaterial as a mechanical network, not only inherits the merits of fast computation and nonvolatile memory from the snap-through transitions of multistable metamaterials, but also features relatively low structural complexity and is less affected by the attenuation during signal propagation. With this, sophisticated high-level learning tasks, such as handwriting recognition[19] and long short-term memory[20], can be performed in a benchmark multistable stacked Miura-origami (SMO) metamaterial[21,22].

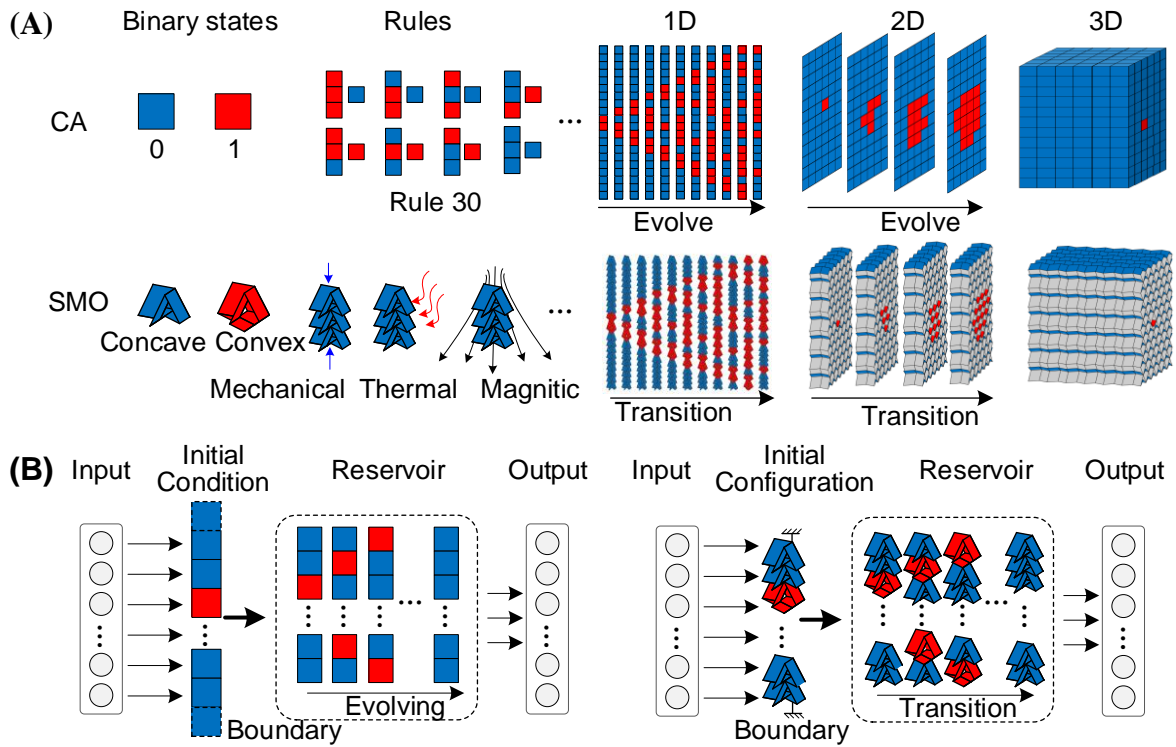

**Figure 1** Conceptual analogy between standard Cellular Automata (CA) and multistable SMO-metamaterials. A) Similarities between CA and SMO metamaterials in terms of the binary states, evolution rules, and transition behavior in 1D, 2D, and 3D space; B) comparison of the computational frameworks based on CA and SMO metamaterials.

The similarities between CA and SMO-metamaterials are manifolds. The first thing to note is that they are both made up of a regular grid of cells, and each cell stays in one of a finite number of states. A typical CA cell is shown in Figure 1A, where two different states are represented by two different colors; similarly, an SMO cell has two geometrically different stable configurations (i.e. concave and convex configurations). Second, the discrete states of individual cells are updated locally and synchronously in accordance with a predetermined evolution rule. For CA, the evolution rule can be a mathematical function (e.g., Rule 30[23]), while for the SMO metamaterial, state transitions can be induced by a variety of different actuation (e.g., mechanical processes, thermal and magnetic fields, etc.) [9,24–26]. Importantly, complex and discriminative transition behaviors are observed in both systems[17,27]. Moreover, they can both be expanded in two or three-dimensional space. As a result, the SMO metamaterial can be considered as a physical analogue of CA, and its computational potential could be unearthed in the context of CA-based computation.

The computational framework for CA and SMO-metamaterials is similar. As discrete Turing universal models, CA have been used to simulate and predict a variety of industrial and biological processes [28,29]. By integrating reservoir computing (RC), a specialized recurrent neural network with pre-determined weights[30–32], into CA, artificial intelligence with learning capability would emerge[33–35]. By feeding the input signal into a CA, rich transitions are generated in much higher dimensional state space by means of pre-defined evolution rules; afterward, the entire dynamic response of the CA is used as a feature vector to be processed in the readout, as shown in the left panel of Figure 1B. Due to its binary nature, CA-based RC is promising for symbolic operations including Boolean logic. Recalling the similarities between CA and SMO metamaterials, we present a physical CA analogue based on the SMO metamaterial to implement RC. In detail, the input is mapped to the initial configuration of the origami cells, and the nonlinear transformation is instead carried out according to a predefined actuation rule. The entire transition sequence of the SMO metamaterial is then fed into a linear readout process (right panel in Figure 1B). We demonstrate that with this computational framework, a simple SMO metamaterial can perform challenging machine-learning tasks (e.g., handwriting recognition and 5-bit memory). The significance of this paper is that we propose a new physical computational framework by establishing an analogy between CA and multistable origami metamaterials in terms of state description and update rules; instead of constructing logic gate networks. This new framework obtains high-level computational power through reservoir computing, enabling more complex learning tasks without adding structural complexity. That is, it does not have the demanding requirements for fabrication of material counterparts and signal propagation when constructing large-scale networks for high-level computation in conventional mechano-logic.

## 2. Results

### 2.1 Training SMO metamaterials for digit recognition

Based on the dual-material 3D printing technique, SMO prototypes are fabricated. Details of the fabrication process are described in Section 6. For a single bistable SMO cell, the experimental force-displacement constitutive profile with two stable states is shown in Figure 2A. The concave and convex stable configurations are respectively interpreted as binary numbers '0' and '1', which are further represented as blue and red squares for visualization purposes. For a 4-layer SMO metamaterial prototype, two connected SMO cells are adopted in each layer to ensure the stability of the structure under axial compression. A series of snap-through transitions occur with quasi-static tensile and compressive loading (Figure 2B). The stable configurations on either side of the unstable switching behavior can be interpreted as 4-bit binary arrays, i.e., 4-pixel rows of blue or red squares. We will subsequently show that these correspondences are used for computation.

To demonstrate the learning ability of SMO metamaterials, we first perform a classification task. The task is to recognize computer-generated images of $5 \times 4$ pixels, see examples in the left panel of Figure 2C, with labels ranging from 0 to 9, respectively. Generally, an SMO metamaterial consisting of $5 \times 4$ cells is needed to encode the entire image. However, for structural simplicity, image segments can be fed into the SMO metamaterial in succession. By encoding the rows containing 4 pixels individually in sequence, the SMO metamaterial prototype only needs to comprise 4 layers.

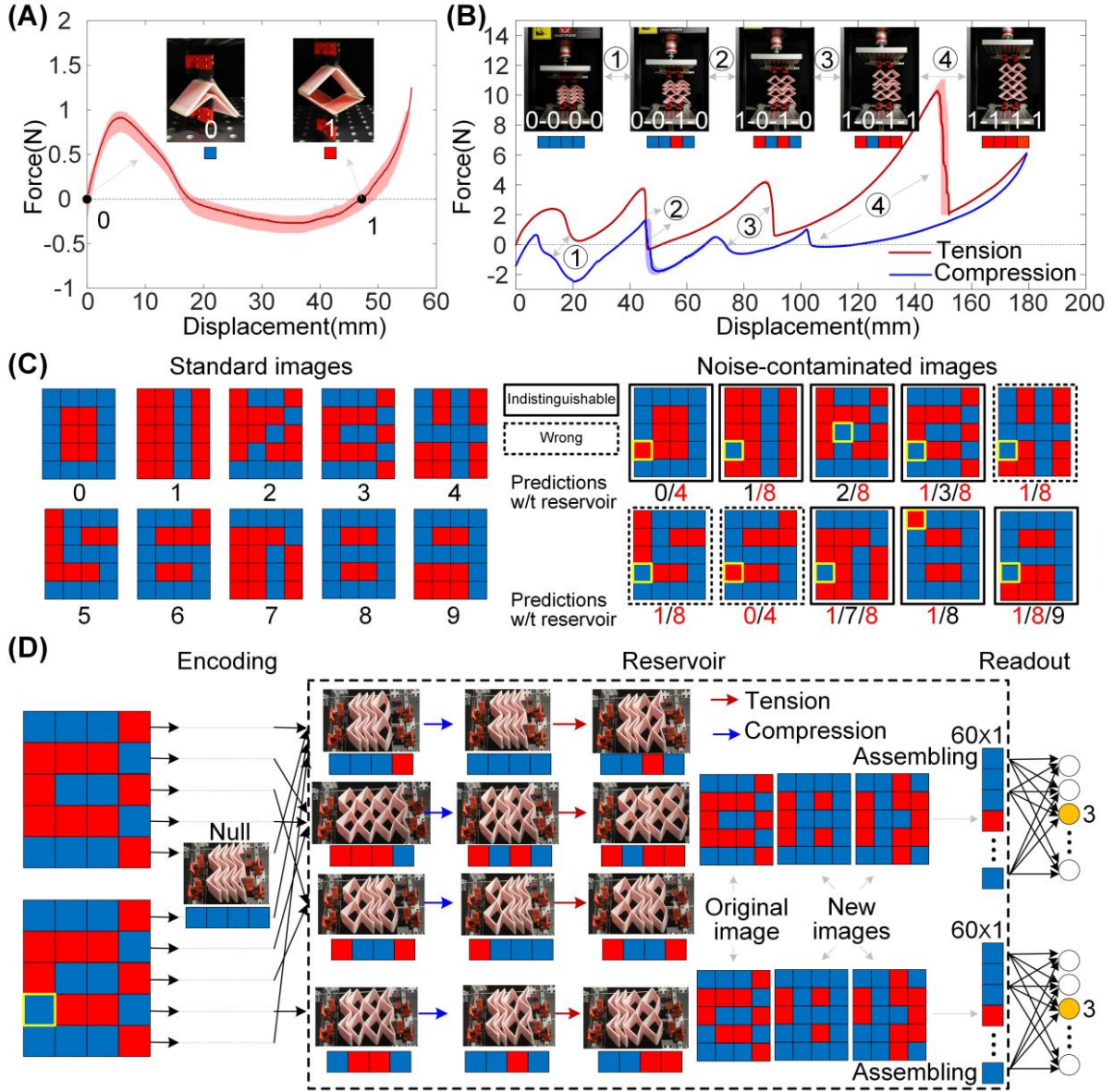

**Figure 2.** Training the SMO metamaterial to recognize digits. A) Force-displacement profile of a single SMO cell, which is obtained by five tensile loading tests; the solid curve and the shaded band denote the average result and the standard deviations, respectively. B) Force-displacement profile of an SMO metamaterial prototype with 4×2 cells; the red and blue curves with shaded bands are the averaged results and the standard deviations obtained under five tensile and compressive tests, respectively; the snap-through configuration transitions are indicated by numbers '1-4'; stable configurations on either side of the transitions and their associated pixels are presented. C) Standard images of digits labeled from '0~9' for training and testing (left); noise-contaminated images for testing (right), the recognized digits without the reservoir are listed underneath, where the numbers marked in black and red indicate the correct and incorrect identifications, respectively. The wrong and indistinguishable predictions are indicated by dashed boxes and empty boxes, respectively. D) The training process to recognize a standard and a noise-contaminated image of digit '3'.

The SMO metamaterial-based computation includes three steps, namely, encoding, transition, and readout. In detail, the binary pixels of each row are first encoded into the SMO metamaterial to generate an initial stable configuration based on the above correspondence, see an example of the digit '3' in Figure 2D. Afterward, a pre-defined actuation rule, i.e., quasi-static tensile and/or compressive loading, is executed in the reservoir to stimulate the transitions. In this specific example, the SMO metamaterial is first compressed to achieve a configuration transition, and then stretched to enable another configuration switch (see Video S1 of the Supporting Information). Hence, for each row of the image (5 rows in total), the corresponding encoded initial configuration is sequentially switched twice following the predefined actuation rule. The dashed box in Figure 2D records and displays the stable configurations of the SMO metamaterial and their associated pixels that are reached sequentially starting from different initial configurations. In addition to the initial stable configurations that correspond to the binary pixels of the original image, two transformed stable configurations are obtained, whose corresponding pixels form two completely new images that contain fresh features helpful for classification. The three images (60 pixels in total) are then assembled and fed into a readout function made up of a neural network with 60 input neurons and 10 output neurons (labeled '0~9'), which outputs the degree of membership of the input image. The output with the highest degree of membership is the predicted digit. To minimize the output error, the readout weights are trained with logistic regression. In this example, the neuron with label '3' gains the maximum output, so the recognized digit is '3'. The other 9 digits in Figure 2A are also recognized experimentally with 100% accuracy under the same reservoir process and the same readout weights.

Another way to further highlight the significance of the SMO metamaterial-based computation is to create a new test image dataset by adding noise to the standard digit figures; samples are shown in the right panel of Figure 2B. Without the transition process of the SMO metamaterial and feeding the noise-contaminated images directly into the readout function, the output is either a completely incorrect prediction (denoted by dashed boxes) or indistinguishable predictions from other digits (denoted by empty boxex), see the right panel of Figure 2C. However, using the transition processes of the SMO metamaterial as a reservoir, all the noise-contaminated images are correctly identified. The function of the transition processes can be viewed as a nonlinear transformation, which can filter out the noisy information and identifies the important characteristics of the original image.

**2.2. Training SMO metamaterials for handwriting recognition**

In addition to the computer-generated images of digits, more realistic and sophisticated images are considered to test the learning ability of the SMO metamaterial-based computational framework. To this end, we use a training and testing database widely employed in the field of machine learning, called the Mixed National Institute of Standards and Technology database (MNIST). The MNIST database is composed of 60,000 training images and 10,000 testing images. To relieve the training burden, we take a reduced set of random portion of these training and testing images, respectively. We will show later that although the reduction of training samples will affect the recognition accuracy, the SMO metamaterial-based computational framework is still effective.

Here, the training and testing of the SMO metamaterials are implemented by simulations. For this purpose, a mechanical model of the SMO metamaterial with an arbitrary number of component cells is developed (see the modeling method and the design parameters of the SMO metamaterial in S1 of the Supporting Information). Based on the principle of least action, state transitions under any predefined actuation rules can be predicted (see the detailed method in S2 of the Supporting Information). To mathematically describe the actuation rule, we define a vector $\chi \in \mathbb{R}^{M \times 1}$, whose elements $\chi_i$ ($i=1,2,\cdots M$) take values of 1 or $-1$, representing the transition under tensile or compressive loading, respectively. The dimension $M$ represents the number of transitions. Based on this model, the learning capability of the SMO metamaterial can be effectively explored, and the readout weights can be trained.

The initial grayscale image of $28 \times 28$ pixels is first pre-processed to binary numbers by transforming the gray pixels into white ones before being used for encoding. Figure 3A shows an example of a handwriting image with a label of '3'. After pre-processing, a $28 \times 28$ binary matrix is obtained. Without loss of generality, each matrix column or row can be interpreted as a 28-element state vector, which can be then encoded into a 28-cell SMO chain to generate an initial stable configuration. The SMO chain is then used to perform a series of transition processes under external loadings. Here, the transitions are implemented by optimization-based simulations. After transitions, a new matrix is obtained by assembling all generated state vectors corresponding to the transformed stable configurations. As an example, the dashed box in Figure 3A exemplifies the transitions of the SMO chain and the images corresponding to the reconstructed matrice. Note that these images are actually distorted versions of the original images. As there are more transitions, the original images get more distorted. Similar to the digit recognition task above, all these matrices corresponding to the distorted images are assembled and fed into the readout, through which the handwriting digits can be recognized. Therefore, the mechanical transition process serves as a nonlinear transformation, with which key features of the original image can be detected, and these features will help in the readout process for classification.

The training and testing processes and results are as follows. 4500 training images and 1000 testing images are randomly selected from MNIST. Configuration switching of the SMO metamaterial chain is executed according to a pre-specified control rule $\chi = \begin{bmatrix} -1 & 1 & -1 \end{bmatrix}$. After three transitions and by training the readout weights, the obtained confusion matrix is demonstrated in Figure 3B. Note that the predicted label from the readout agrees well with the true labels, and the overall recognition accuracy can reach 90.4%, demonstrating the effectiveness of the computational framework. To distinguish the contribution of the SMO metamaterial in computation, simulations without the reservoir (i.e., without the transition processes) are also run as a comparison; in other words, the input images are directly fed into the readout layer, which causes a decrease in recognition accuracy (87.2%). This suggests that the nonlinear transitions of the SMO chain do help distinguish input states. Additionally, this improvement is evident with training data sizes ranging from 3000 to 6000, see Figure 3C.

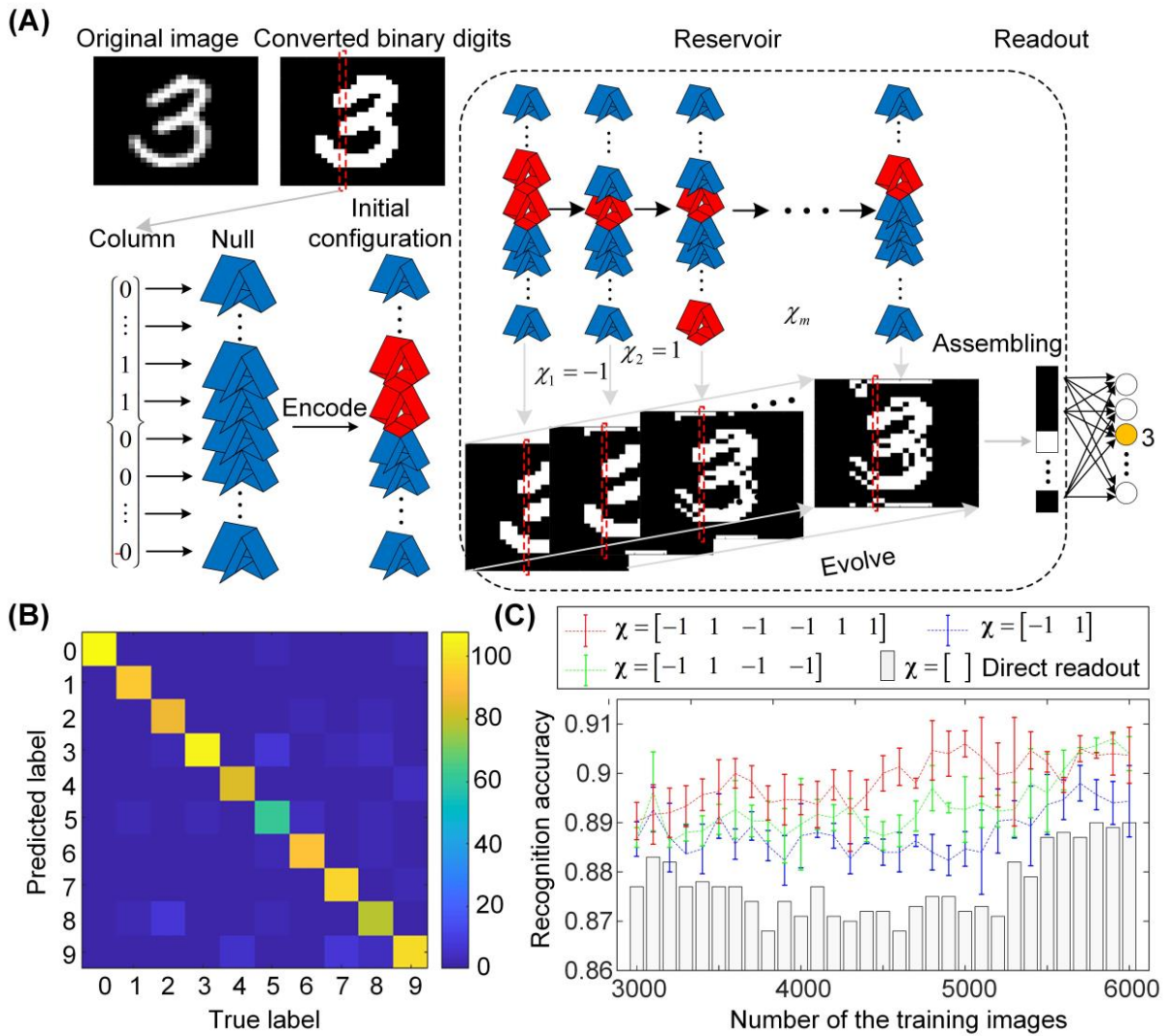

**Figure 3.** Training the SMO metamaterial to recognize handwriting digits. A) Schematic illustration of the training process, including the pre-processing, encoding, transition, and readout. B) The confusion matrix predicted with 4500 randomly-selected training images and 1000 randomly-selected testing images. C) Recognition accuracy corresponding to different actuation rules and different structural properties (five different SMO chains), in which the average values and the standard deviations are plotted.

We further examine the effects of crease stiffness and actuation rules on the computational results of the SMO metamaterial. Five distinct SMO chains with different crease stiffness (detailed in S1 of the Supporting Information) are tested with the same loading process. For different numbers of training images, the averaged recognition accuracy and the associated standard deviations of the five SMO chains are shown in Figure 3C. It reveals that with the same number of training images, more control steps often lead to higher recognition accuracy, with the highest accuracy reaching 91.8%. Such improvement is evident even when SMO cells have different crease stiffness, suggesting that the computation performance of SMO metamaterials is more affected by the actuation rules. However, the structural properties of the SMO cell, which fundamentally determines the transition sequences, are also important in adjusting the recognition accuracy under a fixed actuation rule.

**2.3. Embedding memory in SMO metamaterial-based computational framework**

When it comes to image recognition, different input data are independent of each other, and the output is directly related to the current input; thus, a feedforward neural network learns how to recognize single input data in the context of machine learning[36]. However, considering the way that the human brain learns, experience and historical data are crucial in addressing current events; in other words, sequential learning capabilities are required to handle sequences of data. In machine learning, a dynamic neural network with recursive loops is widely adopted to simulate the historical influence[37]. This inspires us to build a recurrent computational framework with SMO metamaterial by using the final configuration in the previous timestep as the initial configuration of the current timestep for sequential learning. Recall that a bistable cell could store information in nonvolatile memory, the final configuration from the previous timestep carries the historic input information and would contribute to sequential learning.

As a benchmark test of sequential learning, the 5-bit memory problem[20] is examined to exploit the 'memory' capability of the SMO metamaterial (Figure 4A). The 5-bit memory task is defined as follows. There are four binary input channels ($x_1$, $x_2$, $x_3$, $x_4$) and four binary output channels ($y_1$, $y_2$, $y_3$, $y_4$). The first two input channels in the first 5 time steps carry a prescribed pattern with either $x_1$ or $x_2$ set to 1 and the other set to 0. An example is highlighted with shade in Figure 4A. The third input channel serves as a distractor: in the duration period after the first 5 time steps, $x_3$ is set to 1. In this specific case, we set the duration period to be 4 time steps. The fourth input channel carries the cue: when $x_4 = 1$, $x_3 = 0$, and it is the time for the first two outputs to repeat the pattern prescribed in the first five time steps. After the cue, $x_3$ is set to 1 again. The output channels $y_1$, $y_2$ are always 0 before the last 5 time steps, while

$y_3$ remains 1 until the cue is recalled. The fourth output channel $y_4$ is unused and is always set to 0.

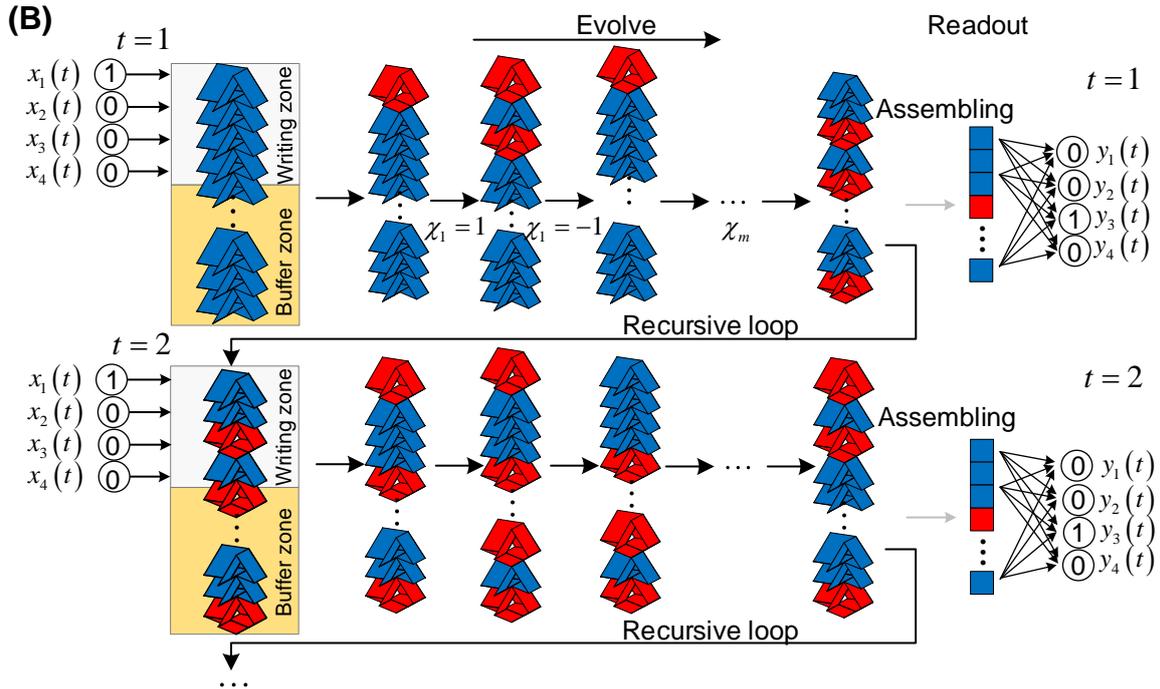

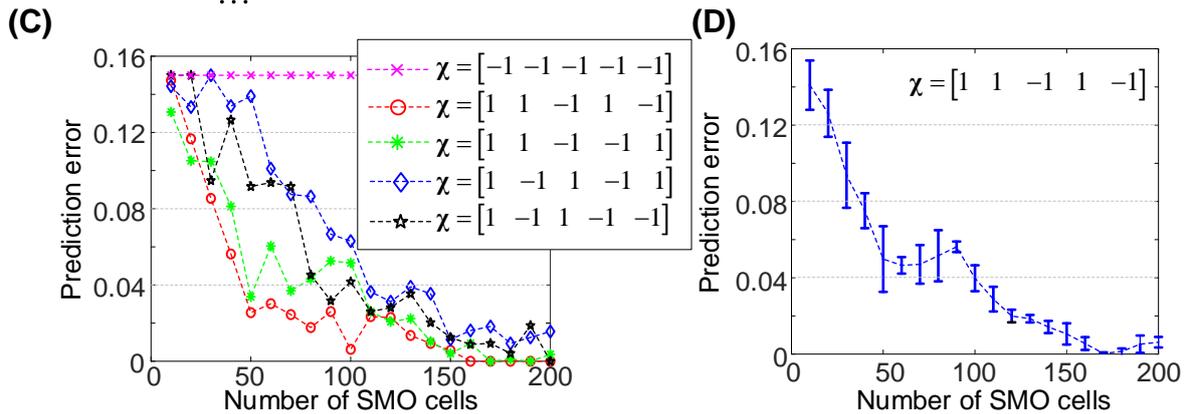

**Figure 4.** Embedding memory into the SMO metamaterial-based computational framework. A) An example of inputs and output for the 5-bit memory task. B) Schematic illustration of the training process of the computational framework in the first and second time step, including encoding, transition, readout, and a recursive loop. C) Prediction errors corresponding to different actuation rules and different numbers of SMO cells. D) Prediction errors corresponding to different numbers of SMO cells and different structural properties (five different SMO chains), in which the average values and the standard deviations are plotted.

Specifically, in this learning task, the SMO metamaterial is asked to 'remember' the input pattern of the first 5 time steps and reproduce it in the last 5 time steps; this is achived by integrating recursive loops in the computational framework. In detail, the constituent cells of the SMO metamaterial are divided into two different zones: the writing or rewriting zone that is responsible for encoding the binary input into the SMO metamaterial, and the buffer or memory zone that consists of more SMO cells than the input dimension to store the historic information (Figure 4B). To implement the 5-bit memory problem with four input channels, four SMO cells are set in the writing/rewriting zone. At every time step, a new binary signal stream is fed into the writing/rewriting zone by encoding the inputs into the four cells, while the configurations of the cells in the buffer zone remain unchanged. After that, the configuration of the SMO metamaterial evolves according to the predefined actuation rules; by recording the entire transition sequences and the corresponding state vectors, an extended vector can be assembled, which is fed into the readout process to get the current output. The final configuration at this time step then serves as the initial configuration of the next time step. At the new time step, the four cells in the writing/rewriting zone are used for encoding the new inputs, while the states of the cells in the buffer zone are kept. The above encoding, transition, readout, and recursion processes are repeated until the last time step, during which, although the information of the first zone is rewritten by new inputs, the historic information is already embodied in the buffer zone through state transitions by predefined actuation rules. Since the states of the cells in the buffer zone will be used as the initial configuration for the state transition at the next time step, this information will play a long-time role, thus endowing the SMO metamaterial with a unique memory function.

The 5-bit memory problem is carried out via the same SMO metamaterial as in Fig. 3. By comparing the output of the SMO metamaterial with the target output, the accuracy of this computational framework in executing the 5-bit memory problem can be determined. Figure 4C displays the prediction accuracy corresponding to metamaterials composed of different numbers of SMO cells and when different actuation rules are applied. Note that with the increase of SMO cells, the prediction error drops accordingly. In the most extreme scenario, if the metamaterial consists of 200 SMO cells, the prediction error gets close to zero. This can be explained by the function of the buffer zone: due to the inclusion of more SMO cells, the prescribed pattern can be retained more in the buffer zone, thus enhancing the memory capability of the SMO metamaterial. The prediction accuracy is also significantly affected by applying different actuation rules (with 6 transitions), especially when the metamaterial is made up of an intermediate number of SMO cells. This is because for metamaterials consisting of a sufficiently few and sufficiently many SMO cells, the prediction results will tend to be stochastic and deterministic, respectively; while for a medium-sized number of SMO cells, an appropriate actuation rule can more effectively embody the predefined pattern in the buffer zone to prevent it from being erased by the new input. Considering an extreme case where an actuation rule of successive compression is applied, the input cannot be effectively transmitted to the buffer zone because the SMO metamaterial will always be transformed into a fully compressed stable configuration no matter what pattern is given; thus, the prediction error is a constant and relatively high regardless of the number of SMO cells included (Fig. 4C).

The variance brought by the uncertainty of the structural propertiesis also investigated. To this end, five SMO metamaterials with different crease stiffness (detailed in S1of the Supporting Information) are generated and used for executing the 5-bit memory task under a given actuation rule, with the prediction errors, in terms of the mean values and the standard deviations, demonstrated in Figure 4D. Overall, both the average prediction error and the variance decrease as the number of SMO cells increases, suggesting that the structural uncertainty is not paramount in the presence of a sufficient number of SMO cells, while structure properties are crucial in the computation when there are not adequate SMO cells.

## 2.4. Implementing computation via a single mechanical actuator

The SMO metamaterial-based computational framework described above can successfully handle complex tasks, but both encoding and configuration transitions require corresponding actuators for implementation, which makes the structure and operation complex. For example, in digital recognition tasks, in addition to the actuator required for configuration transitions, encoding also requires additional actuators, which thus poses challenges for practical applications. To tackle this issue, a default stable configuration of the SMO metamaterial that can be transformed to from any other configurations is specified. For the digit recognition problem demonstrated in Figure 2, the default configuration is '0-0-0-0', which can always be reset with successive compression via a single mechanical actuator. We can then define the initial configuration, which can be any stable configuration reached by prescribed tension or compression processes starting from the default stable configuration (e.g., '1-0-1-0' in Figure 5). Starting from the initial configuration, the SMO metamaterial then experiences reconfiguration under an actuation rule, which is not predefined by users but determined by the input image, with a one-to-one correspondence between the binary numbers and the actuation direction, i.e., '1' and '0' indicate a tension and a compression process, respectively. After the transitions, the SMO metamaterial is reset to its default stable configuration and a new computation is started. The above processes, including encoding the image information, implementing the configuration transition, and resetting, can be accomplished by simply stretching or compressing the entire structure without operating on individual constituent SMO cells, thus reducing the number of required actuators to one.

Through the above improved framework, not only the standard digits but also the noise-contaminated digits in Figure 2C can be successfully recognized with only one actuator (see an example based on the modified framework in Video S2 of the Supporting Information). The modified framework achieves a significant improvement in structural and operational complexity compared to the original framework that requires image information to be written into the SMO metamaterial as an initial configuration. However, any change is a double-edged sword, determining actuation rules from the image information in the computational framework will reduce the flexibility of the transition procedures, thus potentially affecting the adaptability to different tasks.

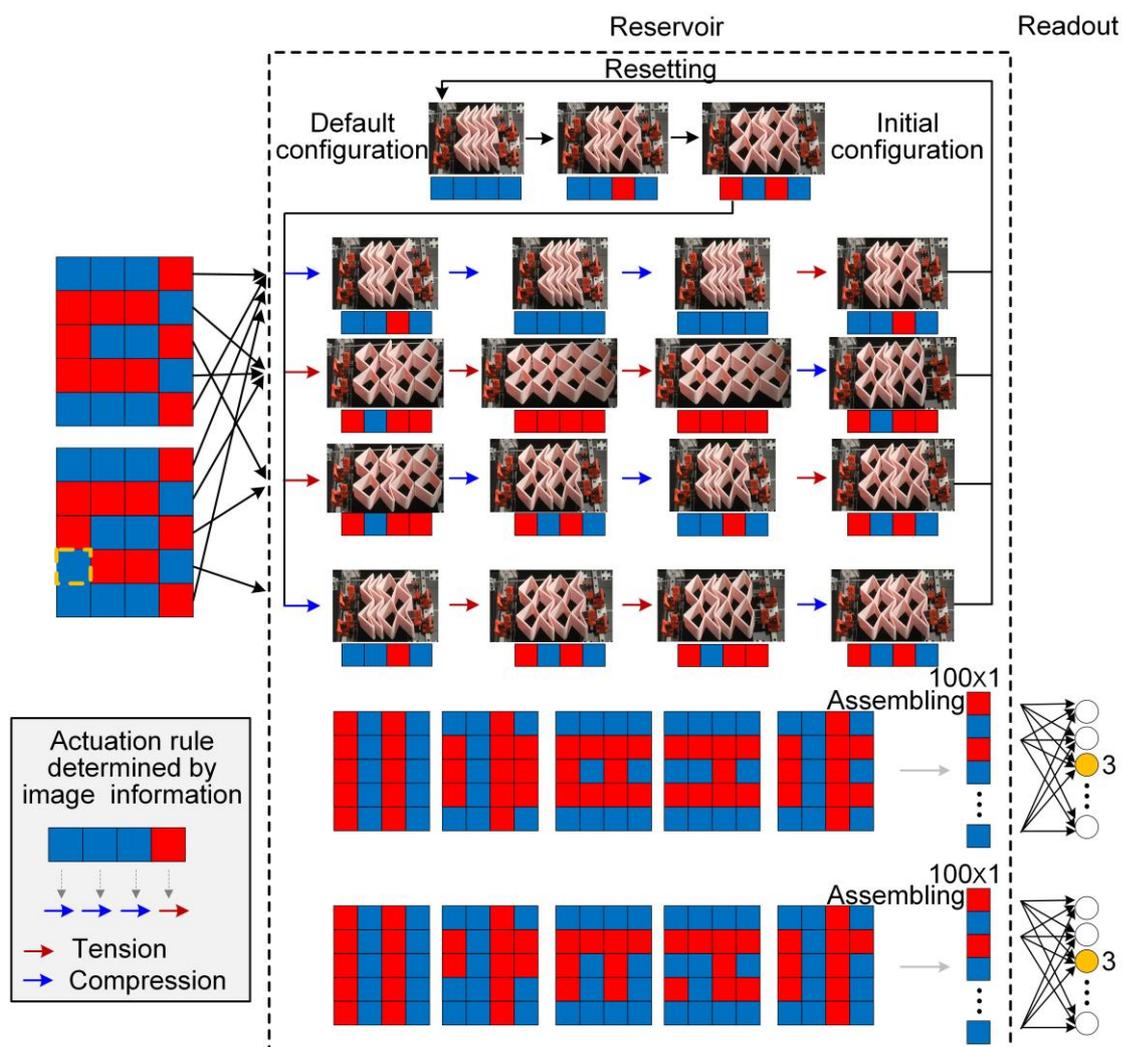

**Figure 5.** The modified SMO metamaterial-based computational framework with only one actuator to encode the image information, implement the configuration transition, and reset. The actuation rule determined by the binary numbers of the input image is demonstrated in the bottom-left inset.

## 3. Conclusion

Multistable origami metamaterials are highly comparable to cellular automata in terms of constituent cells, stable states, transition sequences, and dimension extension capability. This inspired us to create a physical analogue of cellular automata using origami metamaterials, with which the demanding requirements for fabrication of material counterparts and signal propagation when constructing large-scale networks for high-level computation in conventional mechano-logic can be addressed, and complex computational tasks can be realized with simpler structures. Under external stimulation, the analogue displays rich transition sequences, which can serve as a reservoir for functions similar to nonlinear transformation and signal processing. We experimentally illustrate the computational power of the SMO metamaterial with a digit recognition problem. For more complex tasks, such as handwriting recognition and 5-bit memory tasks, the effectiveness is tested via simulations. The proposed framework and the successful completion of tasks based on it offer a unique paradigm for the development of computationally capable metamaterials, opening the door toward mechanical intelligence.

It should be noted that our computational framework is flexible in terms of encoding methods and actuation rules, allowing for a variety of modifications. For example, the actuation rule can be set in accordance with the requirements of different tasks or based on the input information; configuration transitions can also be implemented by alternative methods, such as thermal, pneumatic, and magnetic ones; we can even split the inputs and feed them into multiple multistable origami metamaterials for parallel computation.

We have demonstrated in this paper the feasibility of employing origami metamaterial for high-level computation, while the question of how to improve computational accuracy and efficiency has not been carefully discussed. Note that the existing results have suggested that the learning capability is closely related to structural properties and actuation rules, which motivates us to further optimize these factors in subsequent research to generate origami metamaterials with stronger mechanical intelligence. In addition, as a conceptual framework, many issues including scaled manufacturing and batch reading need to be addressed in the future.

## 4. Experimental Section

The prototype composed of $4 \times 2$ SMO cells is fabricated by the dual-material 3D printing technique based on a commercial multi-material printer (Ultimaker S5), demonstrated in Figure 6. To ensure the rigidity of the facets and the flexibility of the hinge-like creases, materials with different orders of magnitude of elastic modulus, i.e., polylactic acid (Ultimaker PLA, with Young's modulus of $1879 \pm 109$ Mpa) and thermoplastic polyurethane (Ultimaker TPU 95, with Young's modulus of $9.4 \pm 0.3$ Mpa), are used, respectively. The 1-mm-thick polylactic acid facets are wrapped with 0.25-mm-thick thermoplastic polyurethane for a firm bounding. To generate different crease stiffness, the creases in sheets $\alpha$ and $\beta$ are printed with TPU 95 of 0.25 mm and 1.0 mm thickness, respectively. In addition, cracks are purposefully designed in the middle of the creases of sheet $\alpha$ to further reduce the torsional stiffness (Figure 6B). To generate a partially-folded stress-free configuration, we first fold and hold sheet $\beta$ and then heat it to 140°C for 20 seconds. After cooling, sheet $\beta$ becomes stable at a partially-folded configuration with the folding angle approximating to 60° (Figure 6A). Finally, sheets $\alpha$ and $\beta$ are stitched with fishing lines along the corresponding creases to form an entire prototype (Figure 6(C)). In addition to the stable configuration shown in Figure 6C, reconfiguration can be achieved by folding sheet $\alpha$, see a few examples in Figure 6D.

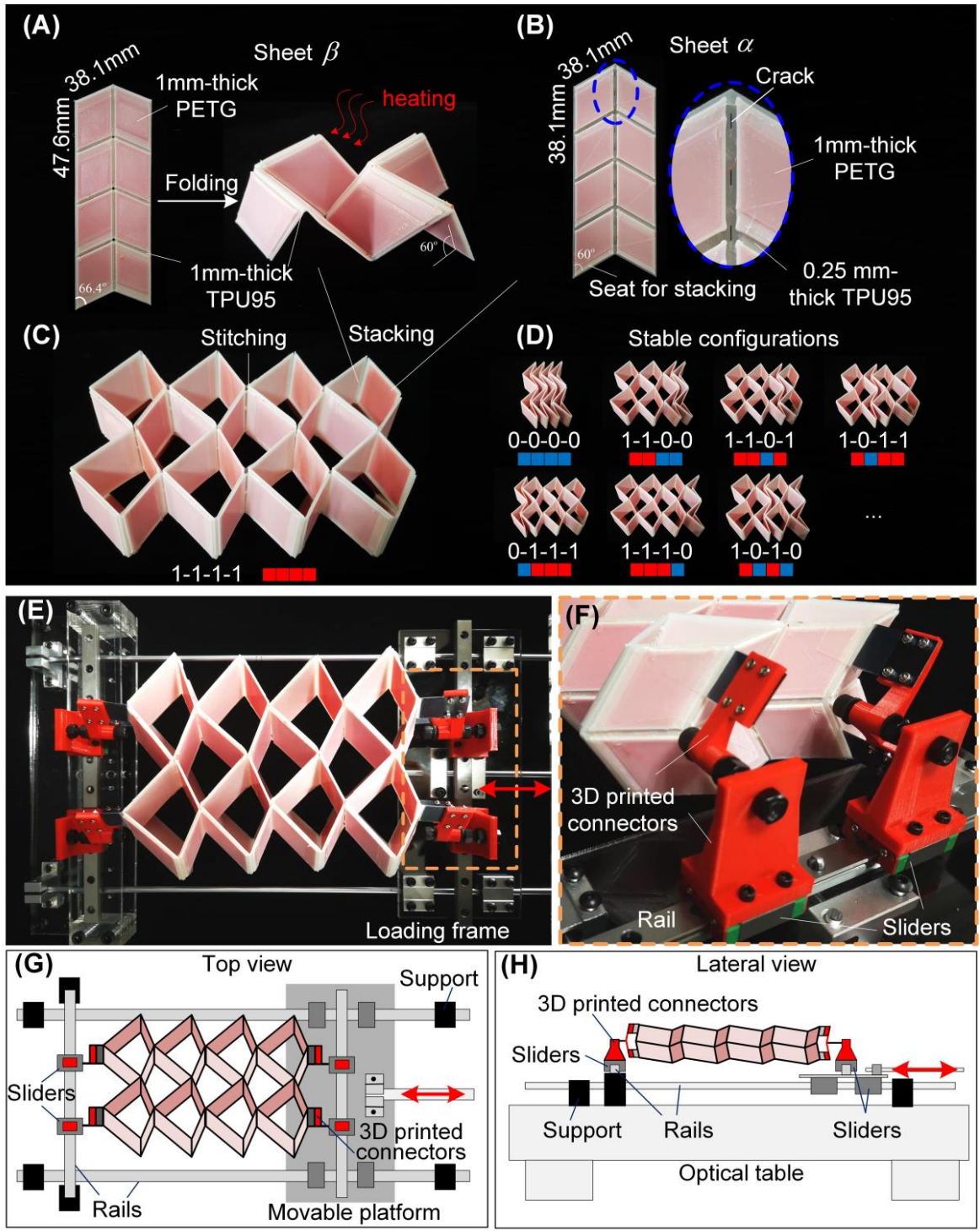

**Figure 6.** The fabrication process of the SMO metamaterial prototype and the experiment setup. (A) Dual-material 3D printed sheet $\beta$ and the heating process to get a partially-folded stable configuration. (B) Dual-material 3D printed sheet $\alpha$ and procedures to reduce the crease stiffness. (C) The experiment prototype obtained by stacking and stitching the sheets. (D) Examples of other stable configurations and their corresponding digits and pixels. (E) and (F) show the photo and zoom-in details of the experiment setup. (G) and (H) display the design of the setup in top and lateral views.

To experimentally transform the configuration of the SMO metamaterial prototype through tension and compression, we specially design a device shown in Figure 6E-H. Specifically, a moveable platform is made for actuation, and several sliders are set on the rail to ensure the free folding of the prototype. 3D printed connectors are used to connect the prototype and the sliders (Figure 6F).

**Acknowledgments**
Z.L., H.F., and J.X. acknowledge the supports from the National Key Research and Development Program of China under Grant No. 2020YFB1312900, the National Natural Science Foundation of China under Grant Nos. 11932015, 12202104, and 12272096, and the Shanghai Pilot Program for Basic Research - Fudan University 21TQ1400100-22TQ009. Z.L. also acknowledges the China Postdoctoral Science Foundation under Grant Nos. 2021TQ0071 and 2021M700819. K.W.W. acknowledge the support of the University of Michigan Collegiate Professorship. The authors would also like to thank Prof. Qi Ge, from Southern University of Science and Technology, for the assistance on dual-mateiral 3D printing.

# Supporting Information

**Cellular automata inspired multistable origami metamaterials for mechanical learning**

*Zuolin Liu, Hongbin Fang\*, Jian Xu, and K.W. Wang*

Zuolin Liu, Hongbin Fang, Jian Xu

Institute of AI and Robotics, State Key Laboratory of Medical Neurobiology, MOE Engineering Research Center of AI & Robotics, Fudan University, Shanghai 200433, China

E-mail: fanghongbin@fudan.edu.cn

Zuolin Liu, K.W. Wang

Department of Mechanical Engineering, University of Michigan, Ann Arbor, MI 48109, USA

## S1. Modeling of the SMO metamaterial

The kinematics of a single SMO cell can be characterized by two different types of parameters. One of them is the geometry parameters of the constituent Miura-ori sheets, i.e., the crease lengths ($a_s$, $b_s$) and the sector angles ($\gamma_s$), where the subscript $s$ denotes the two different Miura-ori sheets ($s = \alpha, \beta$). To ensure stacking, $b_\alpha$ should be equal to $b_\beta$, i.e., $b_\alpha = b_\beta = b$. Besides, kinematic compatibility requires $a_\alpha \cos\gamma_\alpha = a_\beta \cos\gamma_\beta$. The other type of parameter is the folding angles $\theta_s$, which are the dihedral angles between the sheets and the $x-y$ plane. It can be utilized to characterize the folding of a unit cell. For the rigid folding scenario, i.e., the facets are rigid and the creases act like hinges, the SMO unit cell is a single-degree-of-freedom system. Therefore, the folding angle $\theta_\alpha$ can be uniquely used to characterize the folding motion of a unit cell. Other dihedral angles $\rho_i$ ($i = 0, 1, \cdots, 4$) (shown in Figure S1(A)) can be expressed as functions of this independent variable $\theta_\alpha$:

$$\theta_\beta = \arccos\left(\cos\theta_\alpha \tan\gamma_\alpha / \tan\gamma_\beta\right), \rho_0 = \theta_\beta - \theta_\alpha, \rho_1 = 2\arccos\left(\frac{\sin\theta_\beta \cos\gamma_\beta}{\sqrt{1-\sin^2\theta_\beta \sin^2\gamma_\beta}}\right),$$

$$\rho_2 = \pi - 2\theta_\beta, \rho_3 = 2\arccos\left(\frac{\sin\theta_\alpha \cos\gamma_\alpha}{\sqrt{1-\sin^2\theta_\alpha \sin^2\gamma_\alpha}}\right), \rho_4 = \pi - 2\theta_\alpha. \tag{1}$$

Two topological different configurations are identified with $\theta_\alpha < 0$ and $\theta_\beta > 0$ by reassigning the mountain and valley creases of Miura-ori sheet $\alpha$, which are interpreted as '0' and '1' states for an origami mechanical bit, respectively, see Figure S1(B). To distinguish, we present the configurations with different colors.

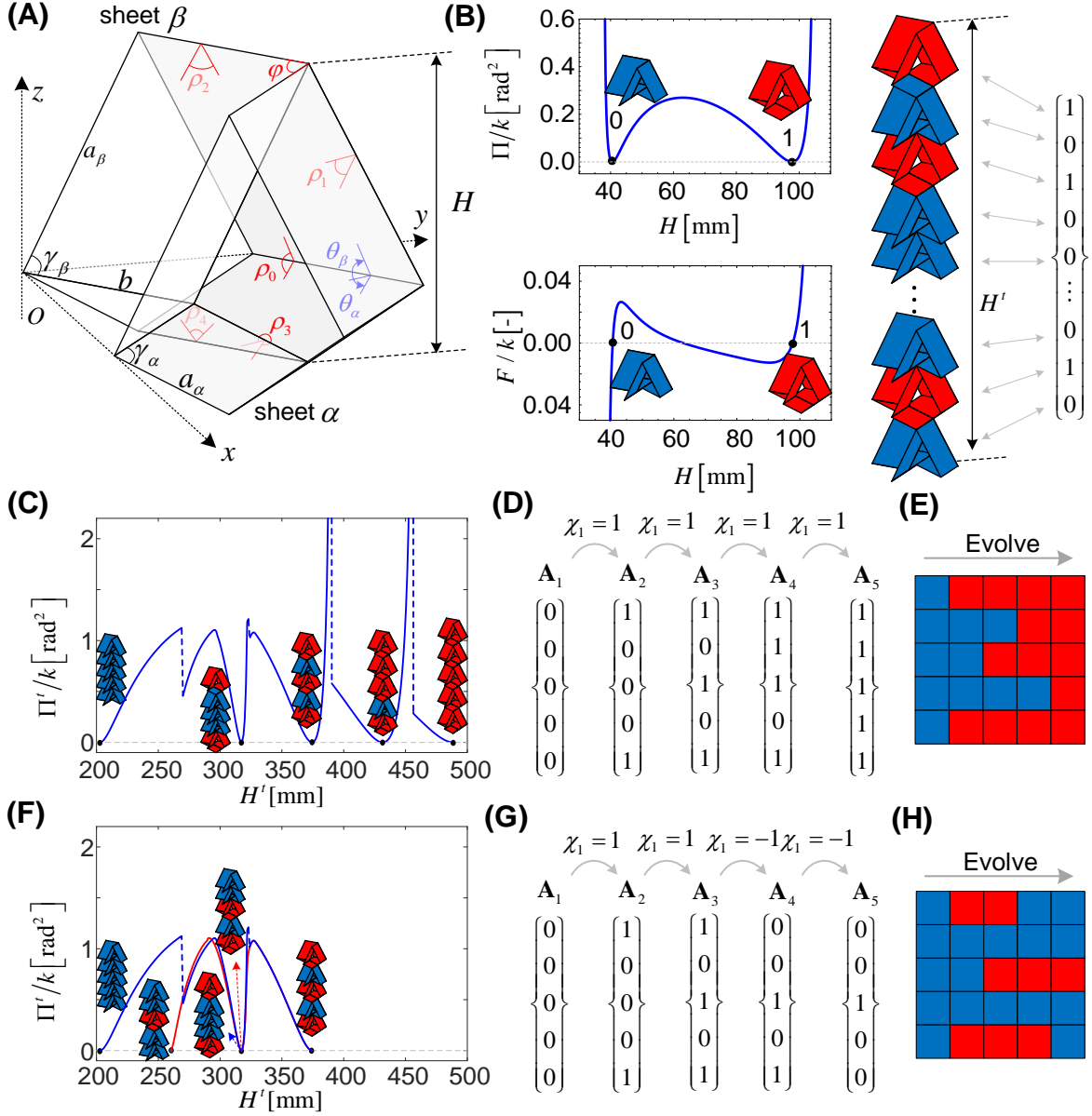

**Figure S1.** Potential energy landscape with respect to the external control and the corresponding mechanical signal streams. (a) and (b) show the potential energy landscape and the mechanical signal streams with an actuation rule $\chi = \begin{bmatrix} 1 & 1 & 1 & 1 \end{bmatrix}^T$; while (c) and (d) present them with an actuation rule $\chi = \begin{bmatrix} 1 & 1 & -1 & -1 \end{bmatrix}^T$.

By assigning torsional stiffness per unit length $k$ for the crease in sheet $\beta$, the elastic potential energy of the unit cell can be written as

$$\Pi = k a_\beta \left( \rho_1 - \rho_1' \right)^2 + k b \left( \rho_2 - \rho_2' \right)^2, \tag{2}$$

where $\rho_1'$ and $\rho_2'$ corresponds to the dihedral angle of the stress-free stable configuration, which can be obtained by substituting the stress-free folding angle (denoted as $\theta'$) to Equation

(1). Since $\rho_1$ and $\rho_2$ are even functions of the folding angle $\theta_\alpha$, the structure can have another stable configuration with $\theta_\alpha = -\theta'$.

The external height of a unit cell can also be expressed as a function of the folding angle $\theta_\alpha$:

$$H = a_\alpha \sin\theta_\alpha \sin\gamma_\alpha - a_\beta \sin\theta_\beta \sin\gamma_\beta. \tag{3}$$

In what follows, the geometry parameters are adopted as

$$b = 38.1\text{mm}, a_\alpha = 38.1\text{mm}, a_\beta = 1.25 a_\alpha, \gamma_\alpha = 60°, \theta' = -\pi/3. \tag{4}$$

The potential energy landscape with respect to the external height present double wells (top panel in Figure S1 (B)), which reals the bistable nature of a single unit cell. The two stable configurations correspond to the two binary states of a mechanical bit, with which information can be stored. By taking the derivative of the potential energy with respect to the external height, the force-displacement relation can be obtained as $F = d\Pi/dH = (d\Pi/d\theta_\alpha)/(dH/d\theta_\alpha)$. We illustrate the force-displacement curve in the bottom panel in Figure S1(C), the two stable configurations are the intersection points with the axis $F = 0$ and with positive slopes. Note that the two states are with different external heights, i.e., the height of state '1' is larger than that of state '0', therefore we can realize the snap-through transition of these two stable configurations by changing the external height, i.e., rewrite the storage of the mechanical bit with height control.

For an origami chain structure with $J$ Miura-ori unit cells connected in series, see the right panel in Figure S1(B), the independent variable $\theta_\alpha$ in $j^{\text{th}}$ unit cell is denoted as $\theta^j$. The vector consisting of all independent variables is denoted as $\boldsymbol{\theta}$. Similarly, the other dihedral angles, elastic potential energy, external height, and the torsional stiffness per unit length are denoted as $\rho_i^j$, $\Pi^j$, $H^j$ and $k^j$ ($i = 0, 1, \cdots, 4$, $j = 1, 2, \cdots, J$), respectively. Besides the elastic potential energy within each unit cell, extra potential energy is induced by the zig-zag connection between the adjacent cells. Define the dihedral angle corresponding to the connecting crease in $j^{\text{th}}$ unit cell as $\varphi^j = 2\tan^{-1}(\cos\theta^j \tan\gamma_1)$, ($j = 2, 3, \cdots, J$), the extra potential energy can be written as

$$\Pi_c^j = \frac{1}{2} k_c^j b \left(\varphi^j - \varphi^{j-1}\right)^2, \tag{5}$$

where $k_c^j$ is the equivalent stiffness per unit length corresponding to the strength of the deformation coupling from the zig-zag connecting creases between cell $j$ and cell $j-1$. Therefore, the total potential energy of the origami chain structure is the summation of the elastic folding energy of every individual unit cell and the potential from their coupling, i.e.,

$$\Pi^t = \sum_{j=1}^{J} \Pi^j + \sum_{j=2}^{J} \Pi_c^j. \tag{6}$$

Similarly, the total external height can be derived as

$$H^t = \sum_{j=1}^{J} H^j, \qquad (7)$$

This Miura-ori chain can be interpreted as $J$ mechanical bits with the one-to-one correspondence between the stable configurations of each individual unit and the binary states. Therefore, a vector $\mathbf{A}$ ( $\mathbf{A} \in \mathbb{R}^{J \times 1}$ ) with $J$ binary elements can be derived to describe the configuration of the chain. For example, the configuration illustrated in the right panel in Figure S1(B) is interpreted as a vector shown on the right. In reverse, a digital vector can also be encoded to a physical configuration of the Miura-ori chain structure.

## S2. Transition sequences with pre-defined actuation rules

For a single-unit cell, the configuration can be uniquely transformed by increasing or decreasing the height. Similarly, the multiple mechanical bits can also be rewritten with the actuation in the height direction. However, the origami chain structure may have multiple feasible configurations corresponding to one total external height. The actual transition depends on the current configuration as well as the loading path. Here, we introduce the optimization strategies to identify the transition sequences.

First of all, we define the actuation rule $\boldsymbol{\chi}$, $\boldsymbol{\chi} \in \mathbb{R}^{M \times 1}$, the $m^{th}$ element of this vector is denoted as $\chi_m$, which takes either 1 ( $\chi_m = 1$ ) or -1 ( $\chi_m = -1$ ) values for increasing or decreasing the total external height of the structure, respectively. $M$ represents the predefined actuation steps. Besides, the folding angles in $m^{th}$ actuation step are denoted as $\boldsymbol{\theta}_m$. Based on the correspondence between the states of the mechanical bit and the stable configurations, the state vector in $m^{th}$ actuation step, denoted as $\mathbf{A}_m$, can be derived with a sign function as

$$\mathbf{A}_m = \left( \mathrm{sign}(\boldsymbol{\theta}_m) + 1 \right) / 2. \qquad (8)$$

Substituting the folding angles $\boldsymbol{\theta}_m$ into Equations (6) and (7), we have the total external height and the total potential energy of the current configuration, i.e., $H_m^t$ and $\Pi_m^t$. Depending on the actuation rule $\chi_m$, we slowly increase or decrease the external height by $\Delta H$ for $\chi_m = 1$ and $\chi_m = -1$, respectively:

$$H_m^t \to H_m^t + \chi_m \cdot \Delta H. \qquad (9)$$

The folding angles corresponding to height $H_m^t$ can be uniquely determined by searching the minimum potential energy in the neighborhood of the current folding angles. The corresponding states vector can be obtained with Equation (8). Once a new stable configuration is transformed, i.e., $\Pi^t = 0$, we update the states vector and start to execute the next actuation rule $\chi_{m+1}$.

We keep the geometry parameters of every single unit cell to be the same (show in Equation (4)). However, in order to introduce the imperfection/asymmetry, we let the torsional stiffness per unit length $k^j$ be a random number around the design value (denoted as $k$), i.e., $k^j = \text{rand}[0.8, 1.2]k$. The equivalent stiffness per unit length is an order magnitude larger than the torsional stiffness per unit length in individual cells and is set as $k_c^j = 50k$. These parameters are used for all the simulations in the main text and the Supporting Information. For different SMO metamaterial, the crese stiffness are all randomly generated by this process.

As a simple case, the total number of the cells is set to be $J = 5$, and the initial configuration of all the unit cells is set at the '0' state. Define the actuation rule as $\chi = \begin{bmatrix} 1 & 1 & 1 & 1 \end{bmatrix}^T$, which represents a successive stretching of the origami chain structure with 4 transitions. The optimized potential energy landscape with respect to the external overall height is presented in Figure S1 (C). We find that the first and the last unit cell switch their configuration from state '0' to state '1' after the first actuation step. Then, the third one, the second one, and at last the fourth one gets transformed in succession with the actuation rules. Very rich transition sequences can be generated by executing this actuation process. From the mechanical memory point of view, the storage of the mechanical bits is rewritten by introducing the actuation rules, and a mechanical signal stream is produced. We demonstrate in the main text that this mechanical signal stream extends the dimensions of the inputs, and can be utilized as computing resources.

In Figure S1 (F), we present the potential energy landscape of the origami chain structure with the same initial configuration but with different actuation rules. In this simulation, after two configuration transitions by increasing the height, the structure is actuated with compression. Note that the paths do not overlap with each other when reversing the loading, instead, new configurations are generated. The corresponding state vectors are presented in Figure S1(G). It can be concluded that with different actuation rules, discriminative mechanical signal streams can be stimulated, which is however crucial for the separation of input states for reservoir computing.

**Movie S1. Experimental Demonstration of Training the SMO Metamaterial to Recognize A Standard Digit 3.**

Experimental demonstration of training the SMO metamaterial to recognize a standard digit 3 in Figure 2(C). Five encoding and transition processes are carried out for training the 5 rows of the image. All the procedures, including the encoding, experimental transitions, and the readout, are visually presented in the movie.

**Movie S2. Experimental Demonstration of Training the SMO Metamaterial to Recognize A Standard Digit 3 with only one actuator.**

Experimental demonstration of training the SMO metamaterial to recognize a standard digit 3 in Figure 2(C) with only one actuator. Resetting and transition processes are carried out for training the image. All the procedures, including resetting, experimental transitions, and the readout, are visually presented in the movie.